\documentclass[onecolumn,showpacs,showkeys,12pt]{revtex4}
\usepackage[cp1251]{inputenc}
\usepackage{graphicx}
\usepackage{dcolumn}
\usepackage{bm}
\usepackage[english]{babel}
\newcommand{\be}{\begin{equation}}
\newcommand{\ee}{\end{equation}}
\newcommand{\bea}{\begin{eqnarray}}
\newcommand{\eea}{\end{eqnarray}}

\begin{document}

\title{The universe evolution as
a possible mechanism \\ of formation of galaxies and their clusters }

\author{
A. Gusev$^1$, P. Flin$^{1,2}$, V. Pervushin$^{1,\star}$\footnotetext{$^{\star}$
E-mail:pervush@thsun1.jinr.ru}, S. Vinitsky$^1$, and A. Zorin$^3$}
\address{$^1$ Bogoliubov Laboratory of Theoretical Physics,
Joint Institute for Nuclear Research, 141980 Dubna, Russia}
\address{$^2$ Pedagogical University, Institute of Physics, Kielce, Poland\\}
\address{$^3$ Faculty of Physics, MSU,  Vorobjovy Gory, Moscow, 119899, RUSSIA}

\date{\today}

\begin{abstract}
 The Kepler problem is considered in a space with
 the Friedmann--Lemaitre--Robertson--Walker metrics of the expanding universe.
 The covariant differential of the Friedmann coordinates (X=a(t)x)
 is considered as a possible mechanism of the formation of galaxies and
  clusters of galaxies.
 The cosmic evolution leads
to decreasing energy of particles, causing free particles to be captured in
bound states. In this approach the evolution of the universe plays the role
usually inscribed to Cold Dark Matter.
\end{abstract}



\maketitle

%\newpage

The description of a Newtonian motion of a galaxy in a gravitational field of
mass of a cluster of galaxies is used for analysis of Cold Dark Matter in the
modern cosmological researches \cite{ch,E:1,E:2,pri}. Here we face with the
following contradiction: the Newtonian motion of a galaxy is described in the
flat space-time $(ds^2)=(dt)^2-\sum_i(dx^i)^2$; whereas the observational data
are analysed in terms of the Friedmann--Lemaitre--Robertson--Walker (FLRW)
metrics \be\label{gr5} (ds^2)=(dt)^2-\sum_ia^2(t)(dx^i)^2. \ee Therefore, it is
worth to study the Newtonian motion of a particle in a gravitational field in
the space-time with the FLRW metrics where observational coordinates of the
expanding Universe are considered as
 \be \label{fc} X^i=a(t)x^i,~~~dX^i=a(t)dx^i+x^ida(t),
 \ee
and instead of the differential of the Euclidean space $dX^i$,
we use the covariant differential of the FLRW space coordinates
\be\label{A0}
a(t)dx^i= d[a(t)x^i]-x^ida(t)=dX^i-X^i\frac{da(t)}{a(t)}.
\ee
Just this problem is considered in this note to study
a possible mechanism of the formation of
galaxies and their clusters, taking into account that both the Newton
motion and the cold dark matter problem should be formulated in  the terms of the
FLRW metrics (\ref{gr5}).

One can check that the interval (\ref{gr5}) in terms of these variables
(\ref{fc}) becomes
 \be\label{A}
 (ds^2)=(dt)^2-\sum_i\left(dX^i-H(t)X^idt\right)^2,
 \ee
where $H(t)=\dot a(t)/a(t)$ is the Hubble parameter.
In the space with the interval (\ref{A}) and the covariant
derivative $(\dot X^i-H(t)X^i)$ the Newton action takes the form
 \be\label{cr11a}
 S_A=\int\limits_{t_I}^{t_0}dt\left[\sum_i\left(P_i(\dot X^i-H(t)X^i)-
 \frac{P_i^2}{2m_I}\right) +\frac{\alpha}{R}\right],
 \ee
 where $\alpha={M_{\rm O} m_I G}$ is a  constant of a Newtonian
 interaction of a galaxy with a mass $m_I$ in a gravitational field
  of a cluster of galaxies of mass ${M_{\rm O}}$.

Let us consider a particle moving in a plane in the cylindrical coordinates
 \be
 \label{coord}
 X^1=R\cos\Theta,~~X^2=R\sin\Theta
 \ee
described by the action  (\ref{cr11a})  in terms of this coordinates
 \be\label{cr11a1}
 %S_A=
 \int\limits_{t_I}^{t_0}dt\left[P_R(\dot R-H(t)R)+P_\Theta\dot \Theta
 -\frac{P_R^2+P_\Theta^2/R^2}{2m_I} +\frac{\alpha}{R}\right].
 \ee
 In this action $P_\Theta=J_I$ is the integral of motion.
 The total energy reads as
 \be\label{def:H}
 E_{\rm tot}(t)=H(t)RP_R+E_N,
 \ee
 where
 $$
 E_N=\frac{P_R^2}{2m_I}+\frac{J^2_I}{2m_IR^2}-\frac{\alpha}{R},
 $$
 is customary Newtonian energy of the system (\ref{cr11a1}).
 The total energy (\ref{def:H}) is not conserved,
 due to the expansion of the Universe, and it  can
 be rewritten in the form
 \bea\label{def:H1}
 E_{\rm tot}(t)&=&\frac{m_I[\dot R^2-H^2(t)R^2]}{2}
   +\frac{J^2_I}{2m_IR^2}-\frac{\alpha}{R}. %\nonumber
\eea

One can see, that in the comparison with the Newtonian energy in the flat space
where $H(t)=0$, an additional term   $H^2(t)R^2$  is appearing. It can be
treated as a friction potential induced by metric (\ref{A}). What is
consequence of this friction?

To know the role of the cosmic evolution of  universe
in a motion of  a particle in the expanding universe,
we consider as an example
a solution of the equation of motion

\be\label{uravnenie}
 \ddot R-{(\dot H(t)+H(t)^2) R}
 -\frac{J^2_I}{m_I^2R^3}+\frac{\alpha}{m_IR^2}=0.
\ee

Let a particle  started
from a point where its total energy is equal to zero
$E_{\rm tot}(t_I)\equiv 0$
 with the initial data ($t=t_I,~R=R_I,~\dot R=0)$.

For simplicity we restrict ourselves by the case of the rigid state \cite{039}
 \be\label{hab1}
 H(t)={H_I\over 1+3H_I (t-t_I)},
 \ee
where $H_I=H(t=t_I)$.

\begin{figure}[ht]
\begin{center}
\includegraphics[width=0.8\textwidth]{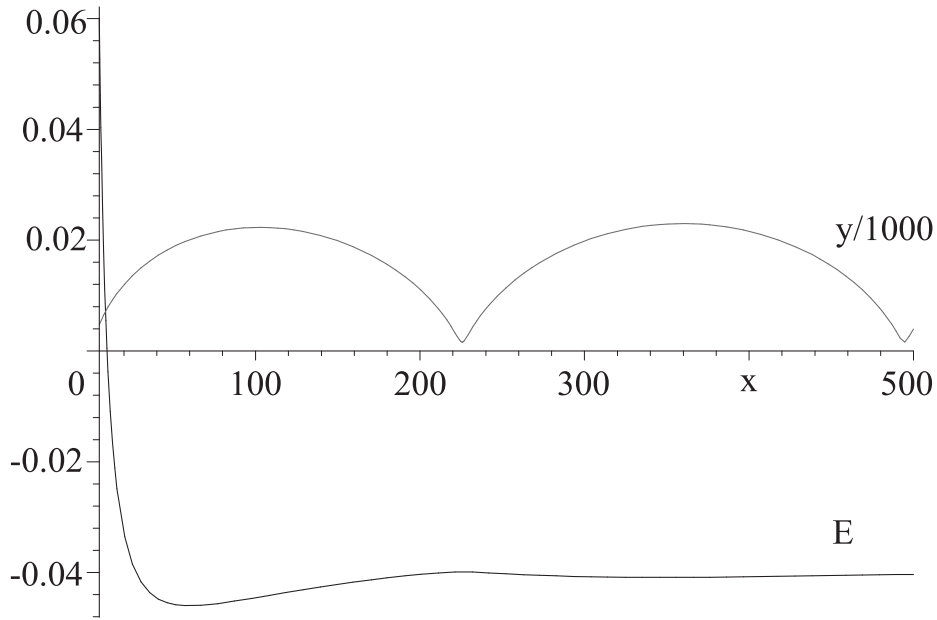}
\end{center}
\caption{\small At the upper panel the numerical solution of the equation
(\ref{cr11a}),
 in dimensionless variables $y(x)=R/R_I$ and $x=H_I(t-t_I)$
 with boundary conditions
$y(x=0)=1$ and  $y'(x=0)=0$ are displayed. The curve at lower panel
demonstrates the
 evolution of the total energy given by (\ref{def:H}).}
\label{fig:h-x}\end{figure}

 Solution of the equations for the case of (\ref{hab1}) in units
 $y=R/R_I,~x=H_I(t-t_I)$ is given in Fig. \ref {fig:h-x} for
 ${\alpha}/{m_IR_I^3H_I^2}=1$, ${J_I^2}/{m_IR_I^4H_I^2}=3$.
 This solution gives a remarkable fact:
later when $H_I(t-t_I) \geq 10$ the energy of this particle becomes negative $E
_ {tot}=E_{tot}/(m_IH_I^2R_I^2) = -0.0405$,
 and   particle is bound.

\begin{figure}[htb]
\begin{center}
\includegraphics[width=0.35\textwidth]{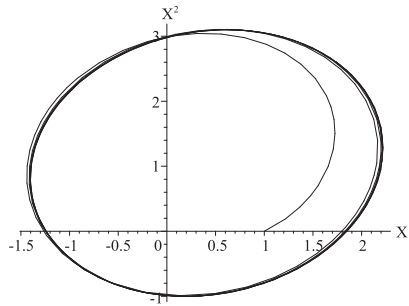}
\end{center}
\caption{\small Pathway of a motion of a particle in the cartesian coordinates
$(X^1,~X^2)$ which  starts at point $(1,0)$ with zero value of the total energy
(\ref{def:H}).} \label{fig:y-phi}\end{figure}

 This solution shows us that the cosmic evolution can form
 the Kepler  bound states such as galaxies and their clusters,
as the cosmic evolution decrease energy of fragments, urging free fragments to
capture in bound states, and free galaxies, in clusters of galaxies. The
lowering of energy which leads to bound state is the influence of friction
appearing in eq. (\ref{A}). Fig. \ref{fig:y-phi} shows us a pathway of a
particle in the cartesian coordinates, that starts at the moment $t_I$ from the
point (1,0) with the zero value of the total energy $E(t_I)=0$.

 It is worth  reminding that the energy conservation law
  $\dot{E}_N(t)=0$ in the conventional Newton theory in the flat space-time
  gives the link of the initial data $v_{I0}$, $R_I=R(t_I)$ at $H=0$

 \be \label{1cr}
 v_{I0}(R_I) =\sqrt{\frac{\alpha}{m_IR_I}}\equiv \sqrt{\frac{r_g}{2R_I}},
 \ee
where $r_g=2\alpha/m_I\simeq 3 \times 10^5 M $ {\rm cm} is the gravitational
radius of an object, and $M$ is
the mass of an object expressed in the solar mass.

 In the considered case of the nonzero Hubble velocity $H\not =0$
 (for week variation of
 the Hubble parameter)  the link of the initial data takes the form
 \be \label{2cr}
 v_I(R_I)=\sqrt{\frac{r_g}{2R_I}+2(H_IR_I)^2}.
 \ee
 From (\ref{2cr}) follows that the Newton theory is not valid for large
 radii
  \be \label{cr}
  R_I\geq R_{\rm cr}=\left(\frac{r_g}{H_I^2}\right)^{1/3}.
  %\simeq10^{20}M^{1/3} {\rm cm} ,
 \ee
 The present-day  value of the Hubble parameter
 $H^{-1}_0\simeq 10^{28} {\rm cm}$
  gives the value of the critical radius
 \be \label{1cr1}
 R_{\rm cr}[M]\simeq 10^{20}M^{1/3} {\rm cm},
 \ee
 where $M$ is the mass of an object expressed in the solar mass.
 One can see that the critical radial distance (\ref{1cr1}) is very close to the
 size of galaxies as well as to galaxy groups and galaxy clusters when adequate
 masses of those structures are taken into account:
 $$R_{\rm cr}[M\simeq  10^{9}]\simeq  10^{23} {\rm cm}\simeq 30 {\rm kpc},$$
 $$R_{\rm cr}[M\simeq  10^{12}]\simeq  10^{24}{\rm cm}\simeq 0.3 {\rm Mpc},$$
 $$R_{\rm cr}[M\simeq  10^{15}]\simeq
   10^{25}{\rm cm}\simeq 3{\rm  Mpc},$$
 and it even coincides with the size of the Coma
 $ R_{\rm size}\sim 3 \cdot 10^{25}cm$ \cite{ch}.

 Let us define the size of a galaxy $R_{\rm size}$ so that
 $\bar v=v_{I0}(R_{\rm size})= v_{I}(R_{\rm size})\sqrt{2/(2+\gamma)}$, where
$\gamma=(R_{\rm size}/R_{\rm cr})^3\leq 1$ (for the  example, if
$R_{\rm size}=R_{\rm cr}/2$, we have $\gamma=1/8$ and
$\bar v=(r_g H_I)^{1/3}\simeq 3 \times 10^8 M^{1/3}$ at $H_I \simeq H_0$).

\begin{figure}[htb!]
\begin{center}
\includegraphics[width=0.6\textwidth]{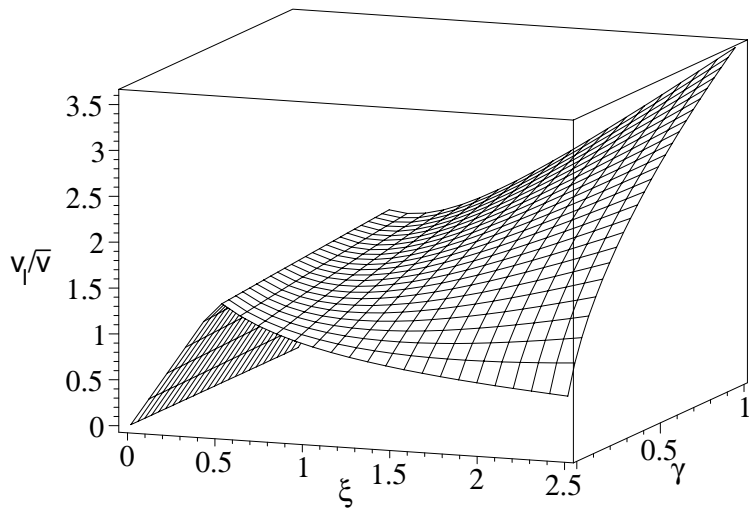}
\end{center}
\vspace{-10mm}
\caption{\small The dependence of circular velocity $v_I/\bar v$ from the
radius $\xi=R/(2R_{\rm size})$ where $R_{\rm size}$
is chosen so that a rotation curve (\ref{2cr}) coincides
with the Newton one (\ref{1cr}):
$\bar v=v_{I0}(R_{\rm size})=\sqrt{2/(2+\gamma)} v_{I}(R_{\rm size})$
at $\xi=0.5$, and
$\gamma=(R_{\rm size}/R_{\rm cr})^3\leq 1$; for $\gamma=1/8$,
$R_{\rm size}=R_{\rm cr}/2$.} \label{halo}
\end{figure}

 Then the rotational curve of the circular velocity -- radius relation
 (\ref{2cr}) can be considered in terms of the ratio $\xi=R/(2R_{\rm size})$ \be
 \label{abc}
 \frac{v_I}{\bar v}=\sqrt{\frac{1}{\xi}+(2\xi)^2\gamma}.
 \ee
 The dependence  (\ref{abc}) of circular velocity $v_I/\bar v$
 from the radius   is
given on Figure \ref{halo}, where the first curve on the
hyper-surface at  $\gamma=0$ corresponds to the Newtonian case, and
  the curves at $\gamma\not =0$ deviate from the
Newtonian case.
Their behaviour imitates
the Cold Dark Matter halos beyond the validity region of the Newton
approximation at $R\geq R_{\rm cr}$ \cite{E:1,E:2,pri}.
  So, the violation of the virial theorem observed in spiral galaxy rotation
curves which is usually considered as an evidence of the existence of the Cold
Dark Matter halos in galaxies, in fact can be interpreted as an evidence of the
Hubble evolution. Moreover we have seen that this cosmic evolution can be
considered as one of the mechanism of galaxies and galaxy clusters formation.

\vspace{8 mm}

{\it One of the authors (V.P.) is  grateful to Professor Richard Swinburne for
calling his attention to problem of Cold Dark Matter to be discussed during the
Notre Dame Conference on Physical Cosmology (30 January -1 February 2003). The
authors thank  Profs. D. Blaschke, M.V. Sazhin, and A.A. Starobinsky and participants
 of the Zelmanov Seminar in ASI for useful discussion of the results
 considered.}

\end{document}